\documentclass[%
 aip,
 amsmath,amssymb,
 reprint,%
]{revtex4-1}

\usepackage{sidecap}
\usepackage{graphicx}
\usepackage{dcolumn}
\usepackage{bm}

\usepackage[utf8]{inputenc}
\usepackage[T1]{fontenc}
\usepackage{mathptmx}
\usepackage{etoolbox}

\makeatletter
\def\@email#1#2{%
 \endgroup
 \patchcmd{\titleblock@produce}
  {\frontmatter@RRAPformat}
  {\frontmatter@RRAPformat{\produce@RRAP{*#1\href{mailto:#2}{#2}}}\frontmatter@RRAPformat}
  {}{}
}%
\makeatother
\begin{document}

\preprint{AIP/123-QED}

\title{The sub-millimetre non-uniformity measurement of residual and coil-generated field in the magnetic shield using atomic vapor cell}

\affiliation{ 
	School of Instrumentation Science and Opto-Electronics Engineering, Beihang University,
	Beijing 100191, China
}%
\author{Chen Liu}
\author{Haifeng Dong$^*$}
 \email{hfdong@buaa.edu.cn}
\author{Junjun Sang}

\date{\today}

\begin{abstract}
Magnetic field source localization and imaging happen at different scales. The sensing baseline ranges from meter scale such as magnetic anomaly detection, centimeter scale such as brain field imaging to nanometer scale such as the imaging of magnetic skyrmion and single cell. Here we show how atomic vapor cell can be used to realize a baseline of 109.6 $\mu$m with a magnetic sensitivity of 10pT/Hz$^{1/2}$@0.6-100Hz and a dynamic range of 2062-4124nT. We use free induction decay (FID) scheme to suppress low-frequency noise and avoid scale factor variation for different domains due to light non-uniformity. The measurement domains are scanned by digital micro-mirror device (DMD). The currents of 22mA, 30mA, 38mA and 44mA are applied in the coils to generate different fields along the pumping axis which are measured respectively by fitting the FID signals of the probe light. The residual fields of every domain are obtained from the intercept of linearly-fitting of the measurement data corresponding to these four currents. The coil-generated fields are calculated by deducting the residual fields from the total fields. The results demonstrate that the hole of shield affects both the residual and the coil-generated field distribution. 
The potential impact of field distribution measurement with an outstanding comprehensive properties of spatial resolution, sensitivity and dynamic range is far-reaching. It could lead to capability of 3D magnetography for small stuffs and/or organs in millimeter or even smaller scale.
\end{abstract}

\maketitle

Measuring the distribution of spin polarization and magnetic field in atomic vapor cell has been pursued for long time. Optical magnetic resonance imaging (OMRI) is used to detect the spin polarization distribution since 1990s\cite{1997Three,1997Optical,1997OpticalRb,1998Alkali}. As a magnetic gradient field for space coding is used in OMRI, the measured spin polarization distribution is usually generated by  pumping light profile instead of field non-uniformity\cite{1999Optical,2000Visualization,2000Diffusion,article,Savukov2015Gradient}. In 2003, photodetector array is firstly used to measure the field distribution in atomic vapor cell with a baseline of 3 mm\cite{kominis2003subfemtotesla}. Since then, this technology has been verified and applied in magnetoencephalography by many groups with a baseline of 3 mm, 4 mm, 5 mm and 7.5 mm, respectively\cite{2009Three, 2018High,Cort2010Magnetoencephalography,2017A}. Charge coupled device (CCD) is also used to record fluorescence emitted from Cs atoms and detect the field distribution in the vapor cell caused by magnetic nanoparticle sample\cite{2016A,2017Characterizing}. In 2017, Shuji Taue et.al. use DMD to measure the radio frequency (RF) field distribution in atomic vapor cell\cite{2020Signal}. While the baseline is sub-millimetre ($\sim$260$\mu$m), the field sensitivity and drift is unknown. Our group measured the spin spatial frequency response of atomic vapor cell using DMD in 2019\cite{2019Dong}. Furthermore, we demonstrate that spin image with a linewidth of 13.7 $\mu$m can be achieved in atomic vapor cell, which is much smaller than the corresponding diffusion crosstalk free distance\cite{2019Spin}. DMD is also used in measuring the coil-generated field distribution under SERF regime, with a sensitivity of $\sim$20fT@30-40Hz and a dynamic range of $\sim$8nT\cite{fang2020high}. 

Here we tried to obtain residual and coil-generated field distribution in the magnetic shield simultaneously with high spatial resolution. We measure non-uniformity of these fields using thermal atomic vapor cell heated to 85 $^\circ$C. Although the effect of shield on the coil coefficient has been realized and measured recently\cite{Ma2021}, we first demonstrate that the distribution of coil-generated field in the shield has strong correlation with the distribution of the residual field due to the hole of the shield.

\begin{figure}[t]
	\includegraphics[width=.5\textwidth]{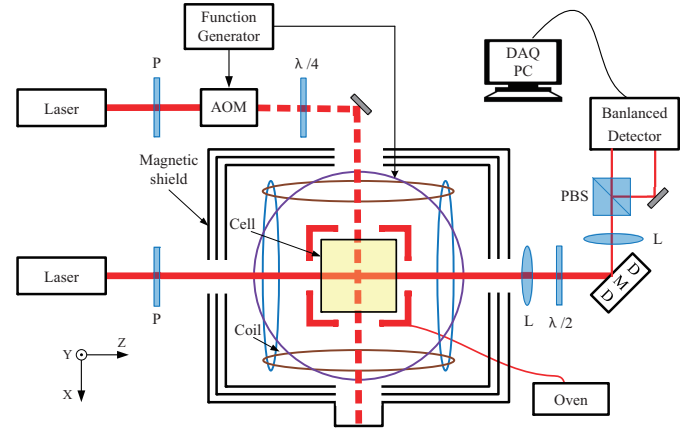}
	\caption{Experimental setup for magnetic field distribution measurement. (P: polarizer; AOM: Acoustic Optical Modulator; $\lambda/4$: quarter-wave plate; $\lambda/2$: half-wave plate; L: lens; PBS: polarization beam splitter; DAQ: Data Acquisition )}
	\label{Experimental_setup}
\end{figure}
The experiment setup is depicted in Fig. \ref{Experimental_setup}. We use orthogonal pump-probe arrangement. Cs atoms are contained in a $\rm 2.5cm \times 2.5cm \times2.5cm $ cubic cell. The cell is filled with 600 Torr of $\rm ^4He$ buffer gas and 150 Torr quenching gas $\rm N_2$. It is placed in an oven heated by non-magnetic twisted wires. We use a set of three nested cylindrical magnetic shields with a shielding factor larger than $10^4$. A set of saddle coil is used to generate radio frequency magnetic field along $y$ direction. Another set of circular coil is used to generate the measured field along $x$ direction. Cs atoms are optically
\begin{figure*}[htb]
	\begin{minipage}{0.25\textwidth}
		\centering
		\includegraphics[width=1\textwidth]{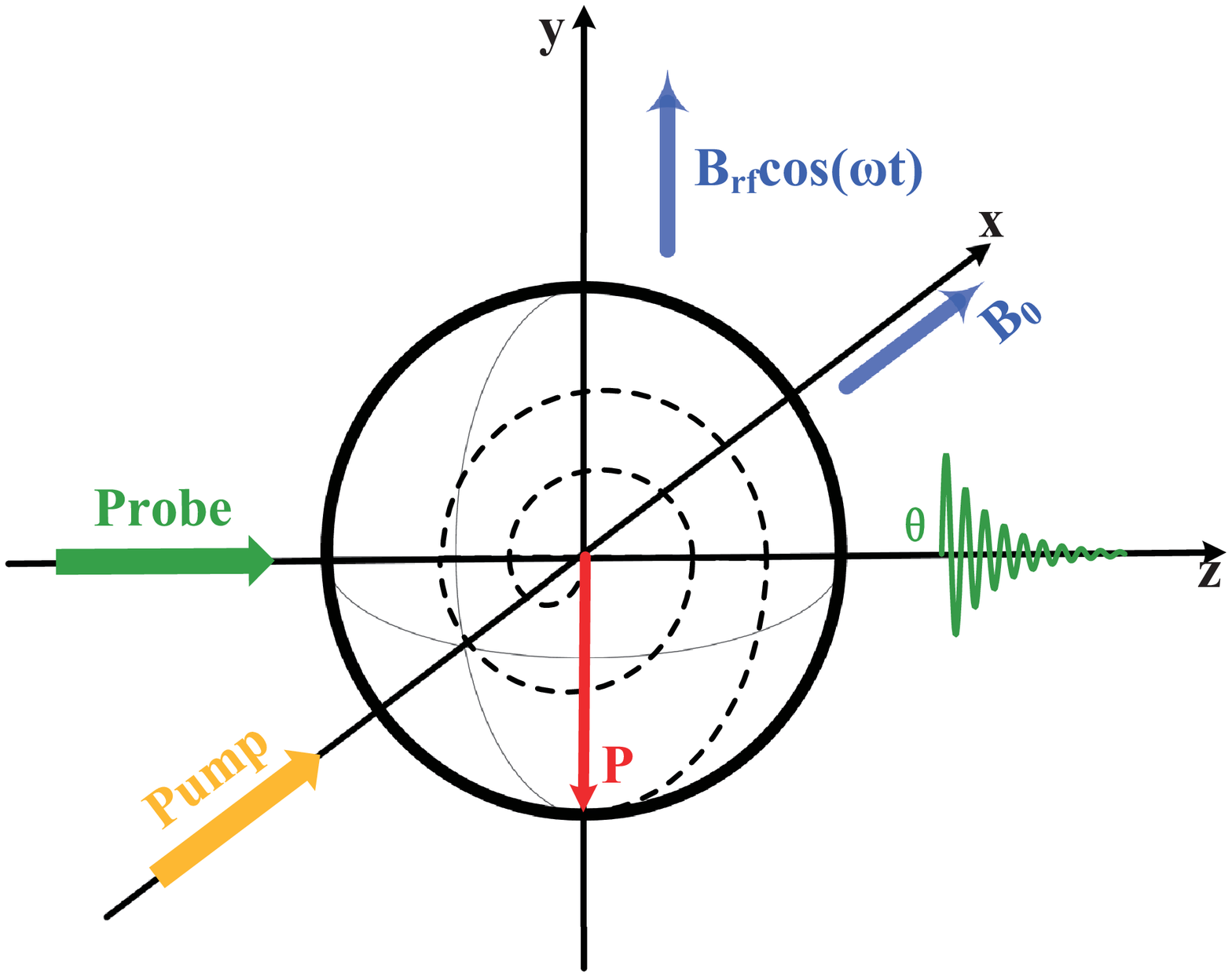}
		\centerline{(a)}
	\end{minipage}
	\begin{minipage}{0.45\textwidth}
		\centering
		\includegraphics[width=1\textwidth]{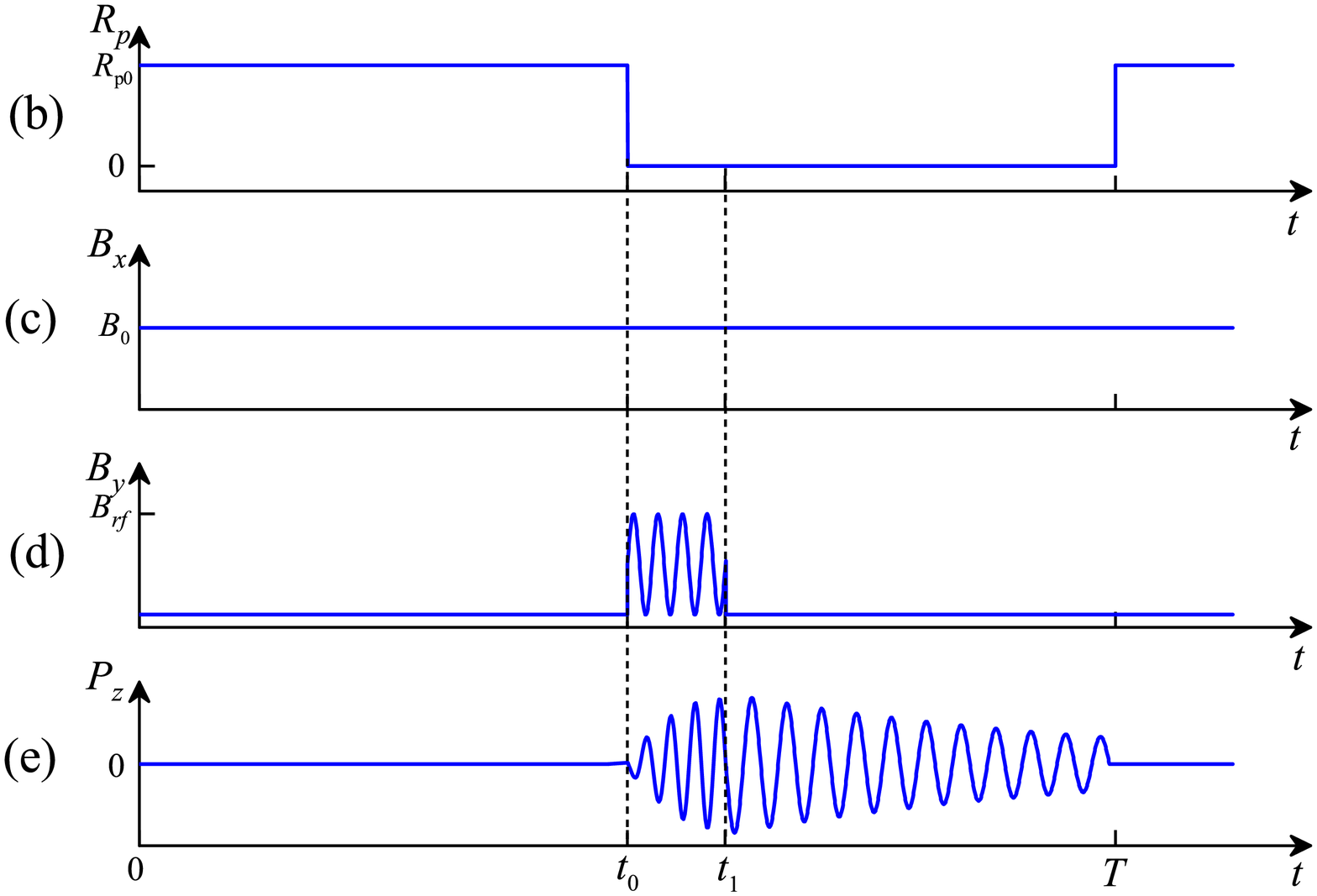}
	\end{minipage}
	\hspace{.3in}
	\begin{minipage}{0.2\textwidth}
	\flushright
	\caption{(a) Schematic diagram. (b)-(e) is the time sequence diagram. (b) The pumping rate along $x$ direction. (c) The magnetic field along $x$ direction. (d) The rf magnetic field along $y$ direction. (e) The change of polarization along $z$ direction over time.}
	\label{ModulatedPumpingSignal}
	\end{minipage}
\end{figure*}
pumped by a 180mW distributed feedback (DFB) diode laser of 894nm (PH895DBR240TS, Photodigm). Another DFB laser of 894nm is used to probe the spin polarization in $z$ direction. The measurement scheme is shown in Fig. \ref{ModulatedPumpingSignal}(a). DMD (Vialux V-7001) is used to scan the spin polarization of different domains. It is comprised of 1024$\times$768 digital micromirrors, each of which is a square reflectors with a side length of 13.7 $\mu$m. We set  8$\times$8 micromirrors as one scanning domain, which corresponds to a baseline of 109.6$\mu$m. Balanced polarimetry is used to measure the rotation of the plane of polarizaiton of probe light corresponding to every selected domain. The differential signal is recorded by data acquisition card NI4461.

To suppress low-frequency noise and avoid scale factor variation for different domains due to light non-uniformity, we use free induction decay to measure the field. Single beam FID magnetometry has been researched in Ref. \onlinecite{2018Free}. There are also double-beam FID schemes where amplitude-modulated pumping light is usually used to polarize the spins and the probe light is used to detect the FID signal\cite{2011An,2021All}. To enhance the signal amplitude, we use RF stimulated FID scheme. Fig. \ref{ModulatedPumpingSignal}(b)-(e) shows the time sequence diagram of the measurement. Cs atoms are pumped from 0 to $ t_0  $ and the pumping light is shut-off then by an acoustic optic modulator (TEM-80-10-894, BRIMROSE). After that RF signal with a frequency of $\omega_0=\gamma B_0$ is turned on from $ t_0  $ to $ t_1  $ so that the spin polarization turns from $x$ to $y$, as show in Fig. \ref{ModulatedPumpingSignal}(a). After that FID spin polarization signal is probed by Faraday rotation of the linearly-polarized probe light. Both the duration $\Delta t=t_1-t_0$ and amplitude of RF are optimized to obtain the maximum FID amplitude.

The evolution of spin polarization is described by the Bloch equation\cite{1946Nuclear}
\begin{equation}
	\frac{\mathrm{d}}{\mathrm{d} t} \mathbf{P}=\gamma \mathbf{P} \times \mathbf{B}+R_{p}(s \hat{z}-\mathbf{P})-R_{r e l} \mathbf{P}
	\label{bloch}
\end{equation}
where $\gamma$ is the gyromagnetic ratio, $R_p$ is the pumping rate, $R_{rel}$ is the relaxation rate. 

From 0 to $t_0$, $B_x=B_0$, $B_y=0$, $B_z=0$ and $R_p=R_{p0}$. The spins are polarized along $x$ direction and the amplitude at $t_0$ is
\begin{equation}
		P_x(t_0)=\dfrac{R_{p0}}{R_{p0}+R_{rel}} 
	\label{SteadyStateSolution}
\end{equation}

From $t_0$ to $t_1$, $R_p=0$, $B_x=B_0$, $B_y=B_{\rm rf} \cos(\omega t)$ and $B_z=0$. The RF magnetic field duration $\Delta t=t_1-t_0=\dfrac{\pi}{2\omega_{\rm rf}}$, where $\omega_{\rm rf}$ is lamor precession frequency of Cs under $B_{\rm rf}$. The spin polarization is rotated from $x$ direction to $z$ direction by the RF field. We set $B_{\rm rf}=1250 {\rm nT}$, which corresponds to a spin polarization loss of less than 10$\%$. The polarization along $z$ direction at the end of $t_1$ is

\begin{equation}
	P_{z}(t_1)=\dfrac{R_{p0}}{R_{p0}+R_{rel}} {\rm e}^{-\dfrac{\pi}{2 \gamma B_{\rm rf}}R_{rel}} 
	\label{rfSolution}
\end{equation}
After that, RF magnetic field is shut off at $t_1$, then the spin polarization precess and decay in the $yOz$ plane according to Eq. (\ref{finalSolution}). 
\begin{equation}
	\begin{cases}
		P_y(t)=P_{z}(t_1) {\rm e}^{-R_{rel}t} \sin(\gamma B_0 t) \\
		P_z(t)=P_{z}(t_1) {\rm e}^{-R_{rel}t} \cos(\gamma B_0 t)
	\end{cases}
	\label{finalSolution}
\end{equation}

\begin{figure}[b]
	\centering
	\includegraphics[width=.5\textwidth]{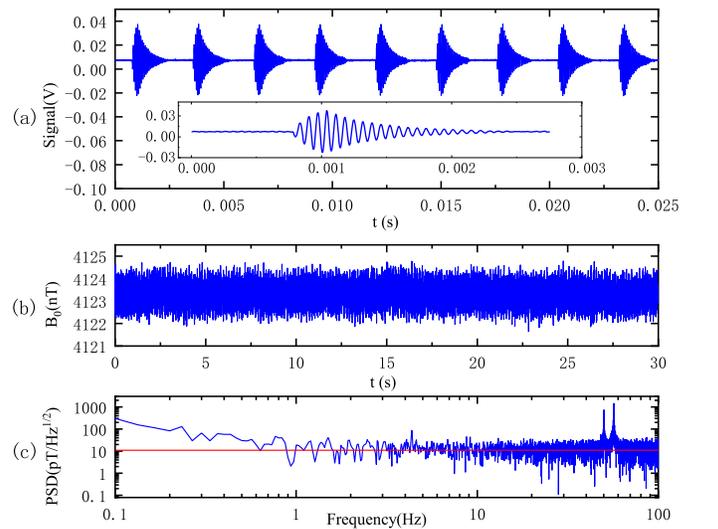}
	\caption{(a) 0.025s FID signal. (b) Magnetic field data obtained by fitting FID signals. (c) PSD of the data in (b).}
	\label{PSD}
\end{figure}
As the signal from balance polarimetry is proportional to $P_z$, we use model Eq. (\ref{finalSolution}) to fit the recorded free induction decay signal.

FID signal of 30s for one domain at current of 44mA is recorded and the data from 0 to 0.025s is shown in Fig. \ref{PSD}(a). After fitting, we obtain about 10710 field measurement values, as shown in Fig. \ref{PSD}(b). Fig. \ref{PSD}(c) is the power spectral density (PSD) which shows a magnetic sensitivity of 10pT/Hz$^{1/2}$@0.6-100Hz. In the scanning measurement, we record five seconds data for every domain under currents of 22, 30, 38 and 44 mA, respectively. The corresponding fields are from 2062 to 4124nT. The total fields of every domain corresponding to these four currents are obtained by FID signal fitting and averaging the five second data.

\begin{figure}[h]
	\centering
	\includegraphics[width=.5\textwidth]{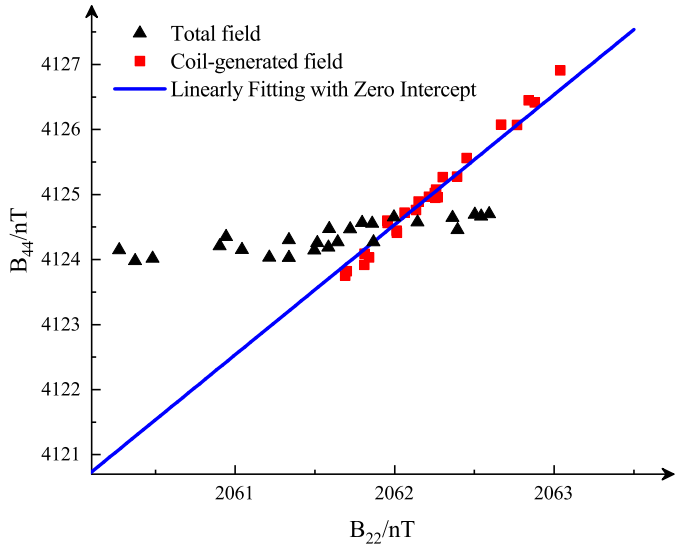}
	\caption{The total fields and the coil-generated fields in different domains corresponding to currents of 22mA and 44mA. The coil-generated fields are obtained by deducting the residual fields  from the total fields.}
	\label{B44_B22}
\end{figure}

By linearly fitting the measured field corresponding to 22, 30, 38 and 44mA at every domain, we obtain the residual field distribution from the intercepts. After deducting the residual fields from total fields for every domains, the coil-generated field distribution is also obtained. To verify whether the residual field is deducted effectively from the total field, we list all the domain's total fields and coil-generated fields corresponding to 22mA and 44mA from small to large In Fig. \ref{B44_B22}. The black triangle represents total fields and the red square represents coil-generated fields which are equal to total fields minus residual fields. The blue dash line in Fig. \ref{B44_B22} shows the fitting result of $ B_{44}=2.00026B_{22} $ with $\rm R^2$=0.9732. The reason why the scale factor is not equal to 2 is that the current accuracy is 0.01mA (Thorlabs LDC 205C). It is clearly seen from Fig. \ref{B44_B22} that the residual field is deducted effectively.

\begin{figure}[!h]
	\includegraphics[width=.5\textwidth]{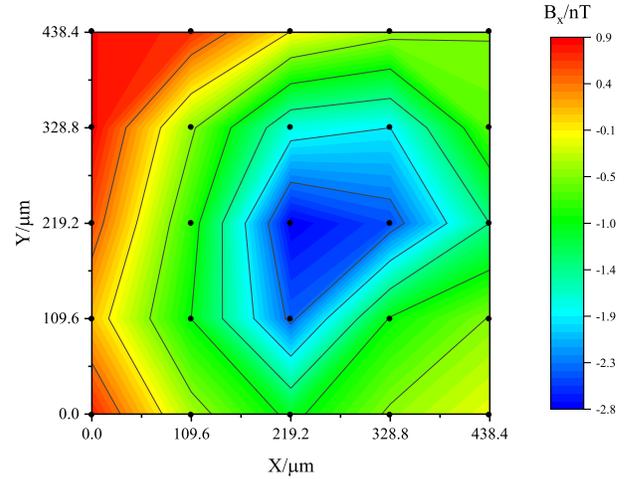}
	\caption{The residual magnetic field distribution in the middle of the shield.}
	\label{res}
\end{figure}

\begin{figure}[hbt]
	\begin{minipage}{0.5\textwidth}
		\centering
		\includegraphics[width=1\textwidth]{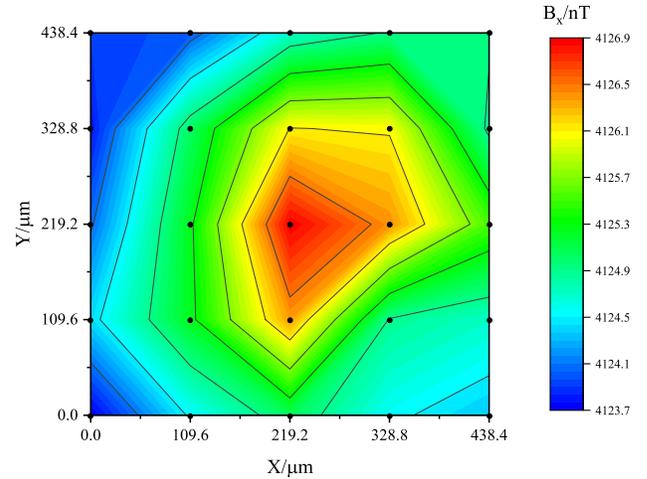}
	\end{minipage}
	\caption{Coil-generated field distribution at a current of 44mA in the middle of the shield.}
	\label{magneticFieldwithoutRes}
\end{figure}

The image of the residual field distribution is shown in Fig. \ref{res} and the image of coil-generated field distribution corresponding to 44mA is shown in Fig. \ref{magneticFieldwithoutRes}. The solid black points in Fig. \ref{res} and Fig. \ref{magneticFieldwithoutRes} indicate the middle positions of the measurement domains.

Comparing Fig. \ref{res} with Fig. \ref{magneticFieldwithoutRes}, we can see that the residual field distribution has strong correlation with coil-generated field distribution. The correlation coefficient is -0.98416. The minus sign means that the hole of the shield has opposite effect on the residual field and coil-generated field distribution. This phenomenon has never been observed. However, it is reasonable considering that the shield can affect the coefficient of the magnetic coil \cite{Ma2021}. So for any experiment and device where the distribution of coil-generated field is important, one should consider the effect of shield structure beyond the Biot-Savart law simulation.

In summary, we have demonstrated a magnetic field distribution measurement scheme with a baseline of 109.6 $\mu$m, sensitivity of 10pT/Hz$^{1/2}$@0.6-100Hz and a dynamic range of 2062-4124nT. 
The residual field distribution and coil-generated field distribution in middle of the magnetic shield are obtained simultaneously.
We found that there are strong correlation between the residual and coil-generated field distribution at sub-millimeter scale. One of the possible reasons is that the hole of the shield affects both of these two field distributions.  
With further improvement of the sensitive domain deployment, this scheme can be used in magnetic field source localization and imaging of millimeter scale objects.


This work is supported by the National Natural Science Foundation of China (61973021).

\textbf{Data Availability} The data that support the findings
of this study are available from the corresponding author
upon reasonable request.

\bibliographystyle{aip}
\bibliography{sample}

\end{document}